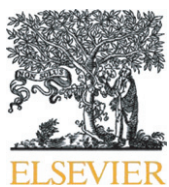
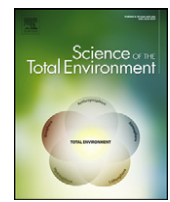

# The effective mitigation of greenhouse gas emissions from rice paddies without compromising yield by early-season drainage

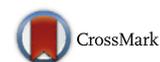

Syed Faiz-ul Islam [a,b,c,*], Jan Willem van Groenigen [a], Lars Stoumann Jensen [b], Bjoern Ole Sander [c], Andreas de Neergaard [b]

[a] Department of Soil Quality, Wageningen University, Droevendaalsesteeg 4, Building 104, 6708 PB Wageningen, The Netherlands
[b] Department of Plant and Environmental Sciences, University of Copenhagen, Thorvaldsensvej 40, DK-1871 Frederiksberg C, Denmark
[c] Climate Change Group, Crop and Environmental Sciences Division, International Rice Research Institute (IRRI), Los Baños, Philippines

## HIGHLIGHTS

- The effects of timing and duration of drainage in rice soils amended with residue were studied.
- Early-season drainage (ED) in combination with midseason drainage reduced $CH_4$ emission up to 90%.
- Yield-scaled GWPs were reduced up to 87% compared to conventional continuous flooding.
- ED results in stabilisation of carbon early in the season, restricting potential for methanogenesis.
- ED is an effective option for small-scale farmers to reduce emissions, water use while maintaining yield.

## GRAPHICAL ABSTRACT

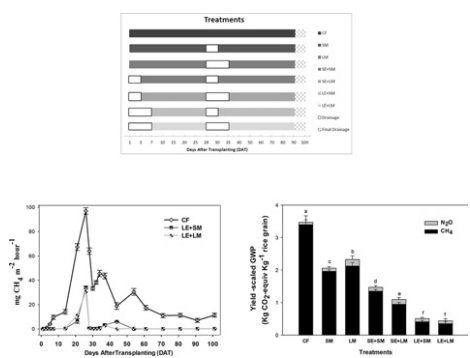

## ABSTRACT

Global rice production systems face two opposing challenges: the need to increase production to accommodate the world's growing population while simultaneously reducing greenhouse gas (GHG) emissions. Adaptations to drainage regimes are one of the most promising options for methane mitigation in rice production. Whereas several studies have focused on mid-season drainage (MD) to mitigate GHG emissions, early-season drainage (ED) varying in timing and duration has not been extensively studied. However, such ED periods could potentially be very effective since initial available C levels (and thereby the potential for methanogenesis) can be very high in paddy systems with rice straw incorporation. This study tested the effectiveness of seven drainage regimes varying in their timing and duration (combinations of ED and MD) to mitigate $CH_4$ and $N_2O$ emissions in a 101-day growth chamber experiment. Emissions were considerably reduced by early-season drainage compared to both conventional continuous flooding (CF) and the MD drainage regime. The results suggest that ED + MD drainage may have the potential to reduce $CH_4$ emissions and yield-scaled GWP by 85–90% compared to CF and by 75–77% compared to MD only. A combination of (short or long) ED drainage and one MD drainage episode was found to be the most effective in mitigating $CH_4$ emissions without negatively affecting yield. In particular, compared with CF, the long early-season drainage treatments LE + SM and LE + LM significantly ($p < 0.01$) decreased yield-scaled GWP by 85% and 87% respectively. This was associated with carbon being stabilised early in the season, thereby reducing available C for methanogenesis. Overall N$_2$O emissions were small and not significantly affected by ED. It is concluded that ED + MD drainage might be an effective low-tech option for small-scale farmers to reduce GHG emissions and save water while maintaining yield.

* Corresponding author at: Department of Soil Quality, Wageningen University, Droevendaalsesteeg 4, Building 104, 6708 PB Wageningen, The Netherlands
E-mail address: syed.islam@wur.nl (S.F. Islam).

## 1. Introduction

Rice (*Oryza sativa*) is the most important agricultural staple for more than half of the world's population and is grown in 114 countries over a total area of around 153 million ha, which is 11% of the world's arable land (FAOSTAT, 2011). Rice production must increase by 40% by the end of 2030 to meet rising demand from a growing world population (FAO, 2009). However, rice cropping systems are considered to be among the major anthropogenic sources of methane (CH$_4$) and nitrous oxide (N$_2$O). Estimates of global CH$_4$ emissions from paddy soils range from 31 to 112 Tg y$^{-1}$, accounting for up to 19% of total emissions, while 11% of global agricultural N$_2$O emissions come from rice fields (US-EPA, 2006; IPCC, 2007). GHG emissions from rice cropping systems have therefore raised serious concerns (van Beek et al., 2010; W. Zhang et al., 2011a; G. Zhang et al., 2011b). There is an urgent need to reduce GHG emissions to the atmosphere to mitigate the adverse impacts of climate change. Therefore rice cropping systems in future will need to combine increased rice yields with decreased GHG emissions.

Methane is the dominant GHG emitted from rice systems in terms of global warming potential (GWP). It is an end product of organic matter decomposition under anaerobic soil conditions (Conrad, 2002; Linquist et al., 2012a, 2012b), therefore the two strategies most often proposed to reduce CH$_4$ emissions are to limit the period of soil submergence (i.e. draining the field) and reduce carbon inputs (through residue management). Of these, water management is the management option most often studied. Several field studies have shown that the drainage of wetland rice once or several times during the growing season effectively reduces CH$_4$ emissions (Wassmann et al., 1993; Yagi et al., 1997; Lu et al., 2000; Towprayoon et al., 2005; Itoh et al., 2011). Through mid-season drainage (MD) and intermittent irrigation, the development of soil reductive conditions can be prevented, leading to reduced CH$_4$ emissions. MD can reduce the total CH$_4$ emission during the rice-growing period by 30.5% (Minamikawa et al., 2014). Additional reported benefits include the reduction of ineffective tillers, the removal of toxic substances and the prevention of root rot, leading to increased yields and reduced water use (Zou et al., 2005; Itoh et al., 2011).

With respect to residue management, according to the International Rice Research Institute (IRRI) about 620 million tons of rice straw are produced annually in Asia alone and this quantity is increasing every year (IRRI, 2016). In most places, these rice straws have no commercial value and are disposed of in various ways. Burning these residues in the field is the most common practice, especially in Asia, because it eliminates numerous pathogens, kills weeds and is less laborious than straw incorporation (Mendoza and Samson, 1999; Kutcher and Malhi, 2010). Both burning in the field (resulting in air pollution and associated health risks) and soil incorporation (resulting in methane emissions) are the cause of environmental concerns. In intensive cropping systems, where two or three crops are grown each year, straw production is so high and the time for residue decomposition so short that farmers traditionally opt to burn the straw. The air pollution associated with this burning is severe because combustion is often incomplete, resulting in the emission of large amounts of pollutants such as SO$_2$ and NOx as well as toxic gases such as carbon monoxide (CO), dioxins and furans, volatile organic compounds (VOC) and carcinogenic polycyclic aromatic hydrocarbons (PAH) (Jenkins et al., 2003; Oanh et al., 2011). Gadde et al. (2009) has calculated that 1 kg of rice straw burnt directly on the field emits 1.46 kg of CO$_2$, 34.7 g of carbon monoxide (CO) and 56 g of dust. The Chinese government and some states in India have already banned straw burning and in the near future many other Asian countries will do likewise because of its association with serious human health hazards, the huge carbon cost (the CO$_2$ price), and loss of essential nutrients such as N, P, K and S (Dobbermann and Fairhurst, 2002; Duong and Yoshiro, 2015). There is therefore likely to be an increased focus on straw incorporation in future, but the application of rice straw usually increases CH$_4$ emissions more than that it counteracts N$_2$O emissions in terms of GWP (Xu et al., 2000; W. Zhang et al., 2011a; G. Zhang et al., 2011b; Xia et al., 2014; Yuan et al., 2014).

However, these studies also suggest that the effects of the straw application on CH$_4$ emissions strongly depend on management (flooding management facilitating aerobic conditions during decomposition) or climatic conditions. The effect of MD on CH$_4$ emissions by altering soil redox conditions is well established, but there is still limited understanding of the effect of the timing and duration of drainage on CH$_4$ emissions. In particular, drainage during the early season (ED) could potentially be effective as the system still contains large amounts of available C from rice straw. ED (as well as MD) could lead to aerobic decomposition and stabilisation of this available C. However, little is known about the effects of ED on either GHG emissions or yield. From the studies of mid-season drainage (MD) and intermittent irrigation, it is evident that although increasing drainage frequency and duration could strengthen the mitigation effect on CH$_4$ emissions (Wassmann et al., 2000), it may have adverse effects on rice plant growth, resulting in a reduction in grain yield (Lu et al., 2000). However, their impacts on rice yield have been found to be mixed, e.g. significantly negative (Towprayoon et al., 2005; Li et al., 2011; Xu et al., 2015), significantly positive (Qin et al., 2010) or unaffected (Minamikawa and Sakai, 2005; Wassmann et al., 2000; Itoh et al., 2011; Yang et al., 2012; Pandey et al., 2014; Vu et al., 2015). The contradictory effects of drainage regimes on rice yield might be attributed to the difference in drainage duration and frequency, the level of water stress during the drainage periods, rice variety and crop management (Belder et al., 2004; Feng et al., 2013). Therefore to ensure food security for the world's increasing population, the impact of any mitigation strategy on GHG emissions and yield should be investigated simultaneously. Moreover the potential trade-offs between CH$_4$ and N$_2$O emissions resulting from MD have been well documented in paddy fields (e.g. Cai et al., 1997; Zou et al., 2005). Johnson-Beebout et al. (2009) have shown in a pot experiment without rice plants that enhanced N$_2$O emissions could potentially outweigh the benefit of reduced CH$_4$ emissions under the alternate wetting and drying (AWD) strategy in terms of GWP. It is not yet clear what the effect of ED would be on N$_2$O emissions. An assessment of the effect of alternative water management strategies on GHG emissions should therefore focus on emissions of N$_2$O as well as CH$_4$.

This study tested for the first time the effect of combinations of ED and MD, varying in timing and duration, on emissions of CH$_4$ and N$_2$O and on yield in a growth chamber experiment with rice straw incorporation. The overall objective was to investigate whether the combination of ED and MD suppresses CH$_4$ emissions from paddy systems incorporated with rice straw while maintaining yield and without correspondingly raising N$_2$O emissions. The specific research questions were: i) Does a combination of ED and MD drainage reduce CH$_4$ emission without significantly increasing N$_2$O emission? ii) Is the length of the drainage period important in terms of GHG emissions? iii) Does early-season drainage affect grain yield and yield-scaled GHG emissions? and iv) What is the relative contribution of different alternative water management regimes to $CH_4$ and $N_2O$ in terms of GWP? The specific hypotheses were: i) in the mid-season drainage treatment, early-season drainage further reduces $CH_4$ emissions without increasing $N_2O$ emissions ii) longer early-season drainage treatments further decrease $CH_4$ emissions but increase $N_2O$ emissions, irrespective of the duration of MD, iii) early-season drainage treatments do not affect grain yield but decrease yield-scaled GHG emissions, and iv) early-season drainage treatments decrease GWP without increasing the relative contribution of $N_2O$ emissions.

## 2. Materials and methods

### 2.1. Experimental site and set-up

The experiment was performed in a growth chamber at the Faculty of Science in the University of Copenhagen, Denmark from April to July 2014. The temperature was 28 °C/22 °C day/night with a 12-h day/12-h night light regime (450 μE m$^{-2}$ s$^{-1}$). The experimental layout was a completely randomised block design with seven treatments and three replicates.

Rice was grown in cylindrical plexiglass soil columns (inner diameter 14 cm, height 30 cm) on which cylindrical headspace chambers (60 cm tall) were mounted for gas sampling and made air-tight with a ring frame with a water seal (Fig. 1). The headspace chambers were sufficiently tall for the Vietnamese early maturing local inbred rice variety (75–85 cm height) used in the experiment. At the bottom of the soil column, a nylon mesh was mounted to enable drainage through a small tube, closed with a simple valve. Two electric fans (4 cm, 12 V DC) were installed inside the gas-sampling chamber, one at the top and the other at the bottom facing opposite directions to ensure an homogenous air mixture. Gas samples were collected through a rubber septum placed in the top of the chamber. To measure the temperature inside the chamber a thermometer probe was inserted through another rubber septum. The air-tightness of the chamber was tested using infrared gas analysis (IRGA) (WMA-2 $CO_2$ analyser, PP Systems, UK) (Ly et al., 2013) and $CO_2$ leakage was found to be minimal (<1%) over a deployment time of two hours.

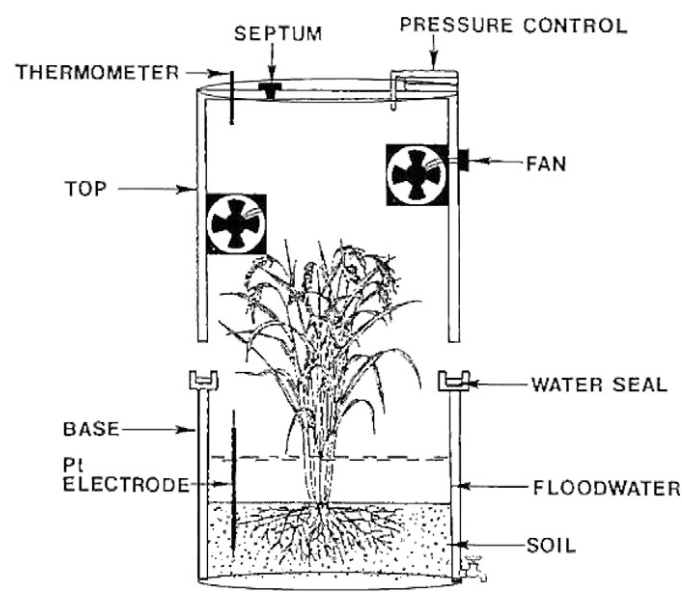

**Fig. 1.** Schematic drawing of soil column and gas chamber (adapted from Lindau et al., 1991) used to collect $CH_4$ and $N_2O$ emissions.

### 2.2. Crop establishment and water regime

Rice seedlings were grown in a nursery bed in the growth chambers for 14 days and then transplanted, with two seedlings per soil column. The soil used for growing rice paddies was sandy loam (Luvisol, FAO/Alfisols, USDA soil Taxonomy), which has highly favourable physical characteristics (high water-holding capacity) for tropical rice production, collected from a NPK-fertilised plot in the Crucial Long-Term Field Trial (Magid et al., 2006). In previous years, approximately 100–150 kg ha$^{-1}$ inorganic N and between 10 and 50 kg ha$^{-1}$ P and K were applied each year. Detailed soil characteristics can be found in Table 1.

The soil was homogenously mixed and 21 columns were filled with an equivalent of 3 kg dry soil each. The water regime varied according to the treatments, which are depicted in Fig. 2. The seven treatments were: i) continuous flooding (CF), ii) short mid-season drainage (SM), iii) long mid-season drainage (LM), iv) short early-season drainage (SE) and short mid-season drainage, v) short early-season drainage and long mid-season drainage (LM), vi) long early-season drainage (LE) and short mid-season drainage and vii) long early-season drainage and long mid-season drainage. After packing the soil, water was added to each column to saturate the soil and then kept saturated for three days. One day before transplanting and before the morning addition of water, basal applications of organic substrates (rice straw) were applied and homogenously mixed with the soil in the columns and then 10 mm water added. On the transplanting day (0 days after transplanting (DAT)), floodwater was drained from the soil columns but kept moist for transplanting. Transplanting was performed early in the early morning (7:00–8:00 am) and after transplanting 10 mm water was added and maintained for 24 h to reduce the transplanting shock for all treatments. From 1 DAT, the soil was kept moist but not flooded (i.e. water table near the soil surface) in all ED (early-season drainage) treatments, while 10 mm standing water above the soil surface was maintained for the rest of the treatments. For the establishment of the seedlings in the ED treatments, a few drops of water were added around the rice shoot twice (morning and afternoon) at 1 DAT. The water level was increased over time during the rice season; the water regime was 10–30 mm for the first 16 DAT, 40 mm for 17–24 DAT and 50 mm from 25 DAT until ten days before harvest. During drainage, floodwater was allowed to evaporate from the soil columns to avoid nutrient loss through drainage and the water maintained at 5–10 mm above the soil surface for just one day before the two periods of fertilizer top dressing. The soil columns were flooded again with demineralised water after fertilisation or drainage periods.

### 2.3. Fertilisation

All treatments received identical applications of mineral fertilizer (N, P and K) and straw. Urea (46% N) was applied as nitrogen fertilizer at the rate of 0.54 g per soil column (corresponding to 160 kg N ha$^{-1}$) in three split doses: 30% of the N fertilizer was applied at 7 DAT, 35% at 35 DAT and subsequently the last 35% at 60 DAT. Single superphosphate (7.2% P) and Muriate of Potash (50% K) were used as P and K fertilisers at the rate of 90 kg ha$^{-1}$ and 60 kg ha$^{-1}$ respectively. Full rates of P and K fertilisers were applied during transplanting. The rice straw applied was collected from Vietnamese rice fields of a local inbred variety. Before application the rice straw was cut into 1–2 cm pieces and then applied at the rate of 62 g per soil column (corresponding to 10 t ha$^{-1}$). The applied rice straw contained 0.6% N, 0.1% P, 0.07% S, 1.7% K, 5% Si and 41% C, and had a C:N ratio of 68.

**Table 1**
Soil properties.

| Properties | Value |
|---|---|
| Soil | |
| Coarse sand (%) | 26 |
| Fine sand (%) | 36 |
| Silt (%) | 17 |
| Clay (%) | 19 |
| pH (1 M KCl) | 6.41 |
| CEC (c mol kg$^{-1}$) | 3.72 |
| Total N (g 100 g$^{-1}$) | 0.18 |
| Total C (g 100 g$^{-1}$) | 2.17 |
| Total P (mg g$^{-1}$) | 0.56 |
| Total K (mg g$^{-1}$) | 173.7 |

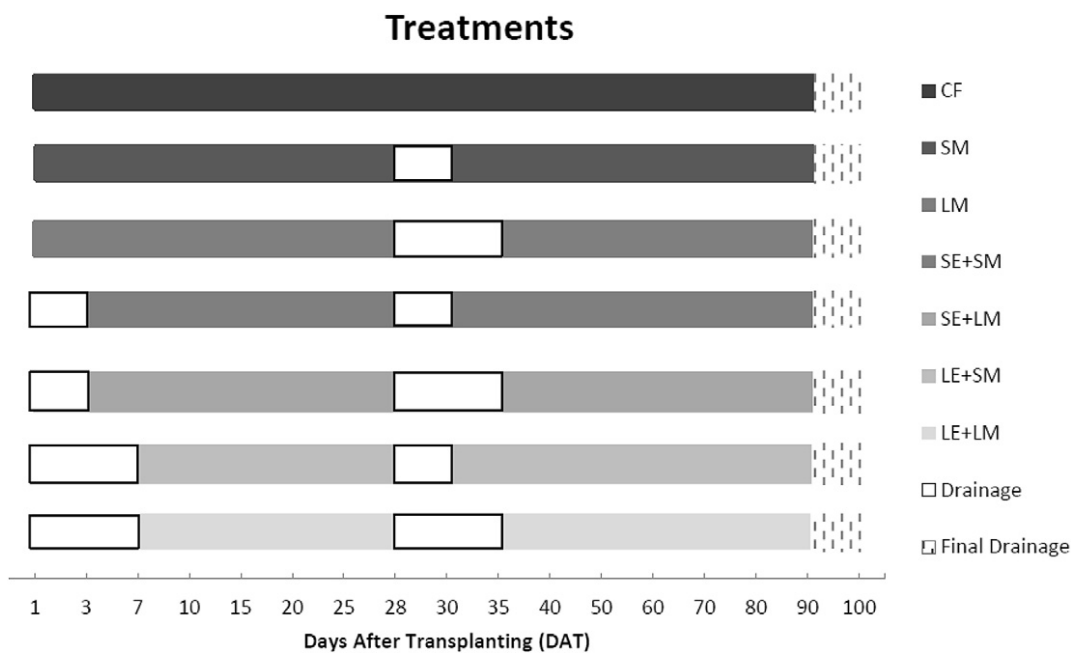

**Fig. 2.** Illustration of treatments CF = *Continuous flooding, SM = Short Mid-season drainage, LM = Long Mid-season drainage, SE = Short early-season drainage, LE = Long early-season drainage.*

### 2.4. Redox and yield measurements

Redox probes (METTLER TOLEDO) were permanently installed in five of the seven treatments, namely CF, LM, SE + LM, LE + SM, LE + LM, because of the limited availability of probes. The redox potential (Eh, mV) readings were taken on each sampling day. At 101 DAT the plant attained physiological maturity with >80% ripe grains. At that point, the aboveground plant biomass and grains were harvested, oven-dried and weighed.

### 2.5. Gas sampling, analysis and calculation

Gas samples were collected on 1, 3, 5, 7, 14, 21, 26, 28, 30, 32, 34, 37, 44, 54, 61, 71, 81, 91, 101 DAT during rice growth, always during the daytime between 8.00 am and 11.30 am. After placing the top chamber on the base, gas samples were taken at 20-min intervals at 0, 20, 40 and 60 min using 10 ml syringes. Gas samples were removed through the rubber septum with a gas-tight syringe and stainless steel hypodermic needle. Collected gas samples were immediately transferred into evacuated 3-ml vial (12.5 mm diameter, Labco limited, UK).

The concentrations of $CH_4$ and $N_2O$ were analysed using a gas chromatograph (Bruker 450-GC 2011) equipped with a separate electron capture detector (operated at 350 °C for $N_2O$ analysis) and flame ionisation detector (operated at 300 °C for $CH_4$ analysis). The oven temperature was set at 50 °C. Helium (99.99%) and Argon + 5% $CH_4$ were used as carrier gases of $CH_4$ and $N_2O$ respectively at a flow rate of 60 ml min$^{-1}$. Certified reference $CH_4$ and $N_2O$ gases were used for calibration and quality control during every batch of gas analyses. The

$CH_4$ and $N_2O$ fluxes were calculated according to Smith and Conen (2004) and Vu et al. (2015).

### 2.6. Global warming potential (GWP)

All emissions were converted to $CO_2$-equivalents (Hou et al., 2012), with GWP for $CH_4$ set at 34 relative to $CO_2$ (based on a 100-year time horizon), and set at 298 for $N_2O$ (IPCC, 2013; Wang et al., 2015). The area-based net GWP of the combined emission of $CH_4$ and $N_2O$ was calculated following Ahmad et al. (2009). Yield-scaled emissions were calculated, as defined by Van Groenigen et al. (2010), as a ratio of growing season net GWP and rice grain yield, expressed in kg $CO_2$-equivalents per kg of grain yield.

### 2.7. Statistical analysis

Statistical analyses were performed using Statistical Analysis System (SAS) Proprietary Software version 9.4 (SAS Institute Inc., Cary, NC, USA). Data were checked for independence, normality and homogeneity of variance. Analysis of variance (ANOVA) with repeated measures on $CH_4$ and $N_2O$ fluxes were performed separately using the Mixed Procedure in SAS. The dependent variables of grain yield, yield components, biomass, seasonal $CH_4$ and $N_2O$ emissions, GWP and yield-scaled GWP were analysed using the GLM Procedure in SAS. The type of response (linear, quadratic or cubic) was identified from the level of significance and an F-test of the difference in $R^2$. All differences were considered to be significant at the 95% level ($P < 0.05$) using Tukey's HSD (honest significant difference) test.

### 3. Results

#### 3.1. Grain yield, aboveground crop biomass

Table 2 shows yield components, grain yield and aboveground biomass. The rice plants in the drainage treatments generally showed more vigour than the plants with CF. The total aboveground biomass was lowest in the CF and highest in the treatment with a long early-season drainage (LED) regime (LE + SM and LE + LM). In contrast, the grain yield was highest in the CF treatment followed by LE + LM and

Table 2
Biomass, grain yield and yield components affected by different alternate water management regimes.

| Irrigation Treatments | Panicles column$^{-1}$ | Spikelets panicle$^{-1}$ | Grain filling (%) | Grain weight (mg) | Grain yield g column$^{-1}$ | Biomass g column$^{-1}$ |
|---|---|---|---|---|---|---|
| CF | 9.3 ± 0.5$^a$ | 104.6 ± 1.5$^a$ | 86.5 ± 1.0$^a$ | 24.2 ± 0.9$^a$ | 20.44 ± 0.3$^a$ | 22.29 ± 0.2$^c$ |
| SM | 9.3 ± 0.8$^a$ | 104.3 ± 1.2$^a$ | 85.8 ± 1.2$^a$ | 23.1 ± 1.0$^a$ | 19.23 ± 0.2$^a$ | 22.89 ± 0.1$^c$ |
| LM | 9.1 ± 0.9$^a$ | 103.1 ± 1.1$^a$ | 84.1 ± 1.3$^a$ | 23.1 ± 1.1$^a$ | 18.22 ± 0.2$^b$ | 23.25 ± 0.1$^b$ |
| SE + SM | 9.3 ± 1.0$^a$ | 104.6 ± 2.1$^a$ | 85.3 ± 1.1$^a$ | 23.3 ± 1.0$^a$ | 19.33 ± 0.2$^a$ | 23.91 ± 0.3$^{ab}$ |
| SE + LM | 9.1 ± 0.9$^a$ | 104.1 ± 1.8$^a$ | 84.6 ± 1.2$^a$ | 23.1 ± 1.2$^a$ | 18.51 ± 0.4$^{ab}$ | 24.01 ± 0.2$^a$ |
| LE + SM | 9.3 ± 1.0$^a$ | 104.3 ± 2.1$^a$ | 86.1 ± 1.2$^a$ | 24.1 ± 1.0$^a$ | 20.16 ± 0.3$^a$ | 24.26 ± 0.2$^a$ |
| LE + LM | 9.3 ± 0.5$^a$ | 104.1 ± 1.8$^a$ | 86.8 ± 1.0$^a$ | 24.3 ± 1.0$^a$ | 20.42 ± 0.3$^a$ | 24.02 ± 0.2$^a$ |

Data shown are the means ± standard deviation of three replicates. Within the column, the values with different letters are significantly different at p < 0.05 level. CF = Continuous flooding, SM = Short Mid-season drainage, LM = Long Mid-season drainage, SE = Short early-season drainage, LE = Long early-season drainage.

LE + SM and the lowest from the LM treatment. However, in terms of grain yield no significant difference was found between the treatments except with LM. Similarly, no significant difference was found in terms of yield components between the treatments.

### 3.2. Methane fluxes

The dynamics pattern of $CH_4$ fluxes over the whole rice-growing period was strongly affected by the water regime (Fig. 3). After the soil columns were waterlogged, $CH_4$ emissions increased steadily until the peak fluxes were attained within the third week after transplanting. The highest $CH_4$ peak was found in the CF treatment at 96 mg m$^{-2}$ h$^{-1}$ at 26 DAT, while the LE + SM and LE + LM treatments involving the LED drainage regime had the lowest peak with 34 mg m$^{-2}$ h$^{-1}$ and 30 mg m$^{-2}$ h$^{-1}$ respectively. Methane fluxes were drastically reduced by the mid-season drainage (SM and LM) period at the end of the tillering stage (from 28 to 35 DAT), which varied between four and seven days based on the treatments. In the late stages of the rice-growing period (54 DAT), small secondary $CH_4$ flux peaks occurred in the CF treatment that were absent from early-season drainage treatments. Fluxes from the CF treatment were consistently higher than those from the other treatments. A one-way ANOVA showed that $CH_4$ emissions from rice plants were significantly affected by water regime and that there was a significant difference between the treatments.

When comparing CF with traditional mid-season drainage varying in length (SM and LM), the peak of SM and LM was observed to be slightly lower than CF (Fig. 3a), after which there was a significant drop in emissions in both SM and LM during the mid-season drainage period. When the short early-season drainage (SED) treatment was included (treatments SE + SM and SE + LM, Fig. 3b), there was an additional decrease in the height of the first peak. This drop was even greater after a longer (7-day) early-season drainage (LED) treatment (LE + SM and LE + LM; graph 3c). Moreover, the post-drainage depression lasted much longer than the mid-season drainage period in the ED treatments depending on the SED/LED treatments.

### 3.3. Nitrous oxide fluxes

$N_2O$ fluxes varied considerably as a result of water regime and time of fertilizer addition in the rice-growing period (Fig. 4). Nitrous oxide fluxes were found to be two orders of magnitude lower than methane fluxes. For all treatments, three $N_2O$ emissions peaks were observed that were all associated with N fertilizer application (7 DAT, 35 DAT and 60 DAT respectively). Treatment CF had the lowest $N_2O$ emissions, whereas the treatment with long mid-season drainage (LM) had the highest $N_2O$ emission. The SED treatments SE + SM and SE + LM showed average $N_2O$ fluxes of 0.25 mg m$^{-2}$ h$^{-}$ and 0.31 mg m$^{-2}$ h$^{-1}$ respectively, about 1.5 and two times higher than that in CF. For the LED treatments LE + SM and LE + LM, the average $N_2O$ fluxes were 0.22 mg m$^{-2}$ h$^{-1}$ and 0.25 mg m$^{-2}$ h$^{-1}$ respectively, about 1.4 and 1.5 times higher than those from the conventional CF treatment. Despite the increase in total drainage duration, the LED treatments LE + SM and LE + LM resulted in reduced $N_2O$ emissions compared to their counterparts SE + SM, SE + LM and LM.

### 3.4. Total cumulative methane and nitrous oxide fluxes

The cumulative $CH_4$ emissions from the paddy soils during the overall rice-growing season showed very significant differences between all the treatments (Fig. 5a). The highest cumulative $CH_4$ emission was recorded from treatment CF with 868 mg per soil column (corresponding to 563 kg $CH_4$ ha$^{-1}$). Long early-season drainage reduced $CH_4$ remarkably, with the LE + SM and LE + LM treatments having the lowest emissions of all the treatments with the cumulative 102 (66 kg ha$^{-1}$) and 88 (57 kg ha$^{-1}$) mg $CH_4$ per soil column respectively. Compared with CF, cumulative $CH_4$ emission was significantly different ($p < 0.01$) and decreased by 88% and 90% in LE + SM and LE + LM respectively. There were no significant differences between LE + SM and LE + LM. However, $CH_4$ emissions from SM and LM were found to be significantly reduced by 53% and 55% respectively compared to CF. SED treatments resulted in 8% and 20% reduction of $CH_4$ emissions when added with the SM and LM treatments, while the LED treatments resulted in a 34% and 35% reduction respectively.

A significant effect ($p < 0.05$) of the water regimes was detected on total cumulative $N_2O$ emissions (Fig. 5b). The lowest total cumulative emission was 2.14 mg per soil column (1.39 kg ha$^{-1}$) from the CF treatment. The highest total cumulative $N_2O$ emissions were observed from LM, followed by SE + LM, with emissions rates of 3.83 (2.49 kg ha$^{-1}$) and 3.77 (2.45 kg ha$^{-1}$) mg per soil column respectively. The total cumulative $N_2O$ emissions from SM and LM were significantly higher ($p < 0.05$) than from CF (Fig. 5b). Similarly, the SED treatments SE + SM and SE + LM were significantly higher (p < 0.05) by 151% and 176% than CF. However, LE showed no significant difference in emission reduction when added to SM, but significantly (p < 0.05) reduced the $N_2O$ emission when added to LM (LE + LM) by 42%.

### 3.5. Redox potential in relation to water drainage

The redox potential (Eh) of the soil during the rice-growing period of the five treatments ranged from 78 to −255 mV for CF, 65 to −246 mV for LM, 72 to −225 mV for SE + LM, 70 mV to −175 mV for LE + SM and 75 mV to −168 for the LE + LM water treatment (Fig. 3d). Generally the redox potential of the soil increased when the soil columns were drained and decreased when the soil columns were re-flooded. Average soil Eh was found to decrease gradually after flooding until the soil columns were drained. After re-flooding, the soil Eh remained negative until the soil columns were drained before harvesting. The redox potential of the LED treatments were significantly different from the CF treatment. In this study, a direct correlation between redox potential and methane flux was found in all the treatments throughout the rice-growing season.

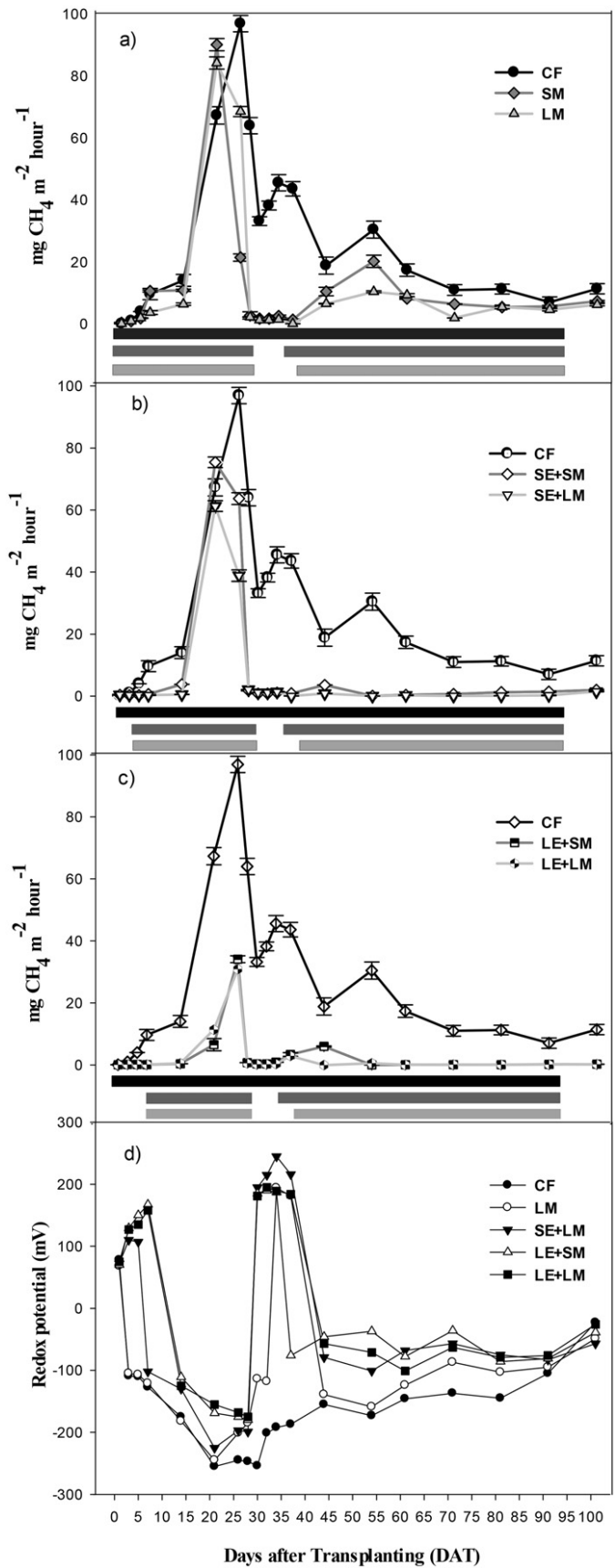

**Fig. 3.** Temporal pattern of $CH_4$ emissions as affected by different alternate water management regimes compared to baseline irrigation regime CF and their relationship with soil redox potential. CF = *Continuous flooding*, SM = *Short Mid-season drainage*, LM = *Long Mid-season drainage*, SE = *Short early-season drainage*, LE = *Long early-season drainage*. Error bars indicate 1 S.E.M. ($n = 3$).

### 3.6. GWP and yield-scaled GWP

There were significant ($p < 0.05$) differences in the impacts of water management on net GWP (Table 3). The highest area-based net GWP ($CH_4 + N_2O$) was 14,507 kg $CO_2$ equiv. $ha^{-1}$ from the CF treatment, which was significantly higher than the other water treatments. Compared with SM and LM, treatments with SED showed a significant reduction in $CH_4$ emission but increased $N_2O$ emissions. However, the LED treatments (LE + SM and LE + LM) resulted in a remarkable reduction of net GWP ($CH_4 + N_2O$) by reducing both $CH_4$ and $N_2O$ emissions compared to SM and LM. Consequently, the significantly lowest net GWP was found in LE + LM (1921 kg $CO_2$ equiv. $ha^{-1}$), followed by the LE + SM (2123 kg $CO_2$ equiv. $ha^{-1}$) and SE + LM water treatments (4228 kg $CO_2$ equiv. $ha^{-1}$).

Similarly, yield-scaled GWP was highest for the CF treatment (3.4 kg $CO_2$ equiv. $kg^{-1}$ rice grain), and significantly ($p < 0.01$) higher than the other water treatments (Fig. 6). Compared with CF, the LED treatments LE + SM and LE + LM significantly ($p < 0.01$) decreased yield-scaled GWP by 85% and 87% respectively. However, no significant difference was found between LE + SM and LE + LM. In contrast, mid-season drainage treatments (SM and LM) reduced yield-scaled GWP on average by just 37%. Moreover, the LED treatments reduced yield-scaled GWP on average by 77% compared to mid-season drainage.

## 4. Discussion

### 4.1. Magnitude and variation of $CH_4$ emissions

The rate of $CH_4$ emission gradually increased with the age of plants during the first three weeks and showed two peaks: one at the early tillering stage at around 21–26 DAT and another at the panicle initiation stage at around 54 DAT in treatments with CF and MD (SM and LM). Treatments with ED (SE + SM, SE + LM, LE + SM and LE + LM) showed only the first peak (Fig. 3). This first emission peak was in line with previous studies, where a first peak was also found between 14 and 30 DAT (Wang et al., 1999; Ly et al., 2013; Vu et al., 2015; Wang et al., 2015). This is predominantly associated with the development of anaerobic soil conditions, readily degradable C from the rice straw amended in the soil, and the rapid growth of rice plants that facilitates the plant-mediated transport of $CH_4$.

For the CF, SM and LM treatments the soil was continuously flooded until 28 DAT, including the first peak. During continuous flooding of paddy soil, trapped $O_2$ is rapidly respired and the soil undergoes reduction processes (Takai and Kamura, 1966). The presence of readily available organic substrates from rice straw in the flooded soil further enhances the reduction process by supplying electron donors, thereby creating an anaerobic environment (Wassman and Aulakh, 2000) leading to methanogenesis (Le Mer and Roger, 2001).

The treatments with ED (SE + SM, SE + LM, LE + SM and LE + LM) underwent a 4–7 day initial drainage period. During this initial period the soil was aerated and Eh increased (between +110 to +167; Fig. 3d). This suppresses methanogenic activity and instead favours methanotrophs to oxidise $CH_4$ (Woese et al., 1978). More importantly it stabilises reactive C (from amendments and soil) by aerobic decomposition, which is likely to progress more quickly during the oxygenated stages, resulting in less substrate for methanogens after subsequent flooding (Pandey et al., 2014). Plant growth was clearly affected by early season drainage, most markedly by larger vegetative biomass. Furthermore, rice plants under aerobic soil conditions have been shown to have less developed aerenchyma compared to those under anaerobic conditions (Kludze et al., 1993), which might have further reduced $CH_4$ transportation and emissions. Therefore in terms of emissions, the difference between treatments with early-season drainage and those with non-early-season drainage is clearly visible due to the delay in the onset of emission and the suppression of the particular $CH_4$ emission peaks that occur early in the cultivation season by ED.

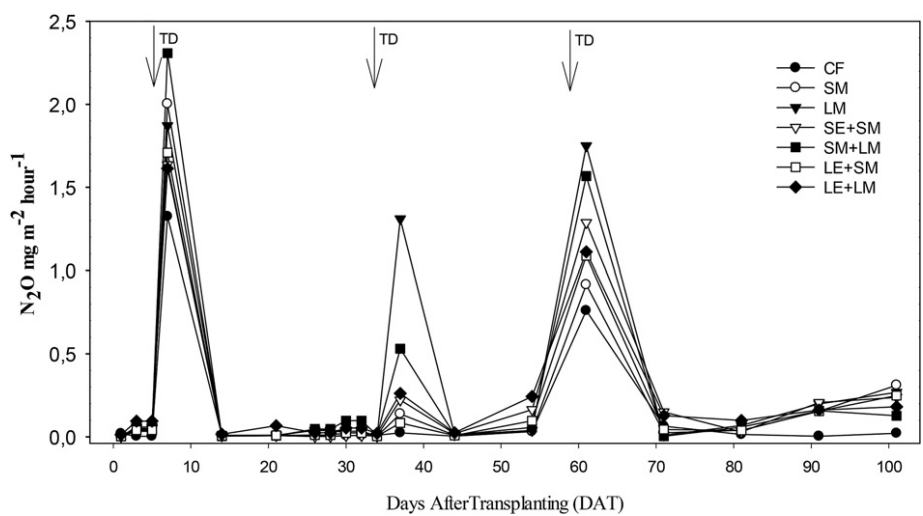

**Fig. 4.** Temporal pattern of $N_2O$ emission as affected by different alternate water management regimes. Error bars are omitted for improved clarity. TD represents top dressing of N fertilizer. CF = Continuous flooding, SM = Short Mid-season drainage, LM = Long Mid-season drainage, SE = Short early-season drainage, LE = Long early-season drainage.

Methane emissions in all treatments peaked in the tillering stage and then decreased gradually due to the complete drainage of the paddy field at 28–35 DAT. According to Zhu (2006), this complete drainage in the later stage of tillering is an important management practice that is used to finish the tiller process of rice and supply rice roots with $O_2$ to prevent sulfide toxicity (Kanno et al., 1997) and also accidentally helps to reduce $CH_4$ emissions. This temporal variability is consistent with patterns observed in other studies of $CH_4$ flux in paddy fields, which have similarly found a decrease in $CH_4$ emissions as a result of mid-season drainage. This could be attributed to the increasingly aerobic conditions in the sediments that suppress methanogenesis and thus $CH_4$ emissions (Jia et al., 2001; Tsuruta, 2002; Towprayoon et al., 2005; Ali et al., 2013; Kim et al., 2013; Singh et al., 2003).

Along with CF, two mid-season drainage treatments (SM and LM) reached a second peak at around 54 DAT. This second $CH_4$ emission peak is in line with previous findings (Schutz et al., 1989; Neue et al., 1997; Ly et al., 2013; Vu et al., 2015). This second peak is attributed to the decay of crop organic matter, such as dead roots and root exudates (Schutz et al., 1989; Neue et al., 1996; Chidthaisong and Watanabe, 1997) during the later stage of rice growth, as well as the slow decomposition of straw under continuous flooding, as reported by most of the authors (Kimura et al., 2004; Gaihre et al., 2011). Furthermore, plant-mediated transport of $CH_4$ is particularly efficient at this stage of plant growth because of the well-developed aerenchymatous system (Wang et al., 2015).

The major difference in flux pattern between the treatments with and without early-season drainage after mid-season drainage was the absence of a second $CH_4$ peak in the early-season drainage treatments. Despite the re-flooding of the paddy pots after mid-season drainage at the final tillering stage, $CH_4$ emissions remained very low, whereas they increased markedly in the non-early season drainage treatments. This could be attributed to further stabilisation of reactive C during the mid-season drainage, which already had a smaller amount of available carbon due to early-season drainage.

The reviews of Minami (1995) and Yan et al. (2009) report that seasonal $CH_4$ emissions range from 2.7 to 1059 kg ha$^{-1}$ for paddy fields around the world. Hence, the seasonal $CH_4$ emissions measured in the present study (57 to 563 kg ha$^{-1}$) are within the range of the published values. The seasonal methane emissions of this study's SED and LED treatments ranged from 57 to 314 kg ha$^{-1}$, which were comparable to recently reported results (18–320 kg ha$^{-1}$) for intermittent irrigation across China (Xie et al., 2010).

### 4.2. Redox potential in relation to water drainage

In this study, higher soil Eh values were observed during the drainage period compared to the soil columns that were continuously flooded. $CH_4$ effluxes increased as the soil Eh decreased, and decreased rapidly after the columns were drained out as soil Eh increased. This is in

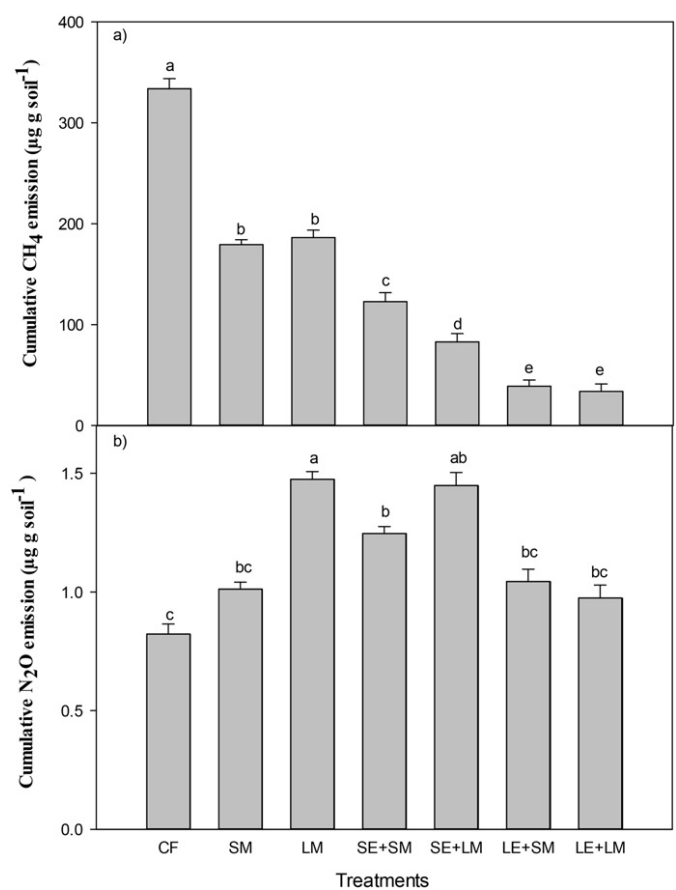

**Fig. 5.** a) Total accumulated $CH_4$ emissions as affected by different alternate water management regimes b) Total accumulated $N_2O$ emissions as affected by different alternate water management regimes. CF = Continuous flooding, SM = Short Mid-season drainage, LM = Long Mid-season drainage, SE = Short early-season drainage, LE = Long early-season drainage. Error bars indicate 1 S.E.M. (n = 3). Different letters indicate significance (p < 0.05) of treatments rice production (small letters).

Table 3
Cumulative emissions of $CH_4$ and $N_2O$ over the WRGS, grain yield and total $CO_2$-e GWP over the 100-year time horizon as affected by organic matter and water management.

| Irrigation treatments | Grain yield (mg g soil$^{-1}$) | $CH_4$ emission (mg g soil$^{-1}$) | $N_2O$ emission (μg g soil$^{-1}$) | GWP $CO_2$-e (mg g soil$^{-1}$) 2007 IPCC, AR4 | GWP $CO_2$-e (mg g soil$^{-1}$) 2013 IPCC, AR5 |
|---|---|---|---|---|---|
| CF | 6.81 ± 0.1[a] | 334 ± 7.9[a] | 0.82 ± 0.05[c] | 8593 ± 212.3[a] | 11,598 ± 283.4[a] |
| SM | 6.40 ± 0.1[a] | 179 ± 2.9[b] | 1.01 ± 0.04[bc] | 4775 ± 84.8[b] | 6385 ± 110.5[b] |
| LM | 6.07 ± 0.2[b] | 186 ± 5.3[c] | 1.74 ± 0.04[a] | 5100 ± 142.8[b] | 6777 ± 190.6[b] |
| SE + SM | 6.45 ± 0.1[a] | 123 ± 7.1[d] | 1.24 ± 0.04[b] | 3441 ± 189.1[c] | 4546 ± 252.9[c] |
| SE + LM | 6.17 ± 0.1[ab] | 83 ± 6.3[e] | 1.44 ± 0.06[ab] | 2504 ± 175.3[d] | 3250 ± 231.9[d] |
| LE + SM | 6.72 ± 0.1[a] | 39 ± 4.1[f] | 1.04 ± 0.06[bc] | 1287 ± 120.5[e] | 1638 ± 157.4[e] |
| LE + LM | 6.81 ± 0.1[a] | 34 ± 5.3[f] | 0.97 ± 0.07[bc] | 1137 ± 152.9[e] | 1443 ± 200.6[e] |

Data shown are means ± standard deviation of three replicates. Within the column, the values with different letters are significantly different at p < 0.05 level. The IPCC GWP factors (mass basis) for $CH_4$ are 25(AR4)/34(AR5) and for $N_2O$ it's 298 times higher than $CO_2$ over the 100-year time horizon, respectively.

agreement with previous studies (Wang et al., 1993; Minami, 1994; Tyagi et al., 2010). According to Masscheleyn et al. (1993) methane is usually formed only after the soil Eh is reduced to below −100 mV with a near neutral pH of the flooded soil. However, other studies have differentiated between critical soil Eh values for laboratory versus field studies and indicated critical redox potentials of −150 mV and −100 mV respectively (Wang et al., 1993; Hou et al., 2000; Tyagi et al., 2010). The onset of significant $CH_4$ emissions was found to start when the soil redox potential dropped below −127 mV for the CF treatment. Overall, the critical Eh value for significant $CH_4$ emissions ranged between −155 to −245 mV for the different water treatments, which is consistent with previous results for laboratory experiments by Tyagi et al. (2010). However, during the post drainage period in the ED treatments, soil Eh was found to pass the theoretical threshold of −100 mV, yet the emissions were not significantly different from zero. This may indicate the absence of C substrate in the system.

4.3. Trade-offs with $N_2O$ emissions

$N_2O$ emissions were low compared to $CH_4$ emissions in terms of GWP. This is in line with previous studies (Abao et al., 2000; Ly et al., 2013; Vu et al., 2015). Bronson et al. (1997) have reported that $N_2O$ emissions are rarely detected during the rice season except directly after fertilisation. Yao et al. (2012) have reported negligible $N_2O$ emissions during three continuous years under flooded conditions. Significant $N_2O$ peaks were evident in all the treatments only after N fertilizer application. This is in line with previous studies (Pathak et al., 2002; Zou et al., 2005; Pandey et al., 2014). Readily available N substrate after topdressing of mineral N might have enhanced nitrification in aerobic zones and subsequent denitrification in anaerobic zones of the rhizosphere, resulting in induced $N_2O$ emissions (Pandey et al., 2014). Based on the literature, fluxes were measured directly after fertilisation, as the peaks were associated with fertilisation and slight changes in sampling day or time of day would have significant effects on the total cumulative emission. In future studies, higher temporal resolution of sampling after fertilisation will benefit the validity of the commutative flux. Furthermore sampling was not undertaken every day, so a small number of measurements account for almost the entire seasonal cumulative flux sampling. There is therefore a possibility that $N_2O$ fluxes were underestimated and the cumulative $N_2O$ flux should be interpreted with care. However, the low variation in the present data suggests that some general conclusions can be drawn. The lowest and highest $N_2O$ emissions were observed in the CF and LM treatments respectively. Prolonged flooding promotes the development of strong anaerobic conditions in soils, reducing any $N_2O$ produced in the paddy fields to $N_2$ (Ussiri and Lal, 2013). Long mid-season drainage could create prolonged partly anaerobic conditions in the soil, favouring simultaneous nitrification and denitrification and resulting in considerable fluxes of $N_2O$ (Davidson et al., 2000; Pathak et al., 2002). These findings are in line with greenhouse experiments by Johnson-Beebout et al. (2009), who report that alternate wetting and drying increases $N_2O$ emissions from paddy soils relative to a continuously flooded treatment. Some field studies have also found that MD and intermittent irrigation increase nitrous oxide ($N_2O$) emissions compared to the CF treatment (e.g. Yan et al., 2000; Nishimura et al., 2004; Towprayoon et al., 2005; Jiao et al., 2006; Zou et al., 2005, 2009). However, $N_2O$ emissions accounted for <15% of the relative total annual emissions and did not eliminate the overall reduction in global warming potential (Tsuruta et al., 1998; Kurosawa et al., 2007; LaHuea et al., 2016). Despite the higher total soil aeration duration in LED treatments, these treatments showed the potential to emit less $N_2O$ compared to the MD treatments SM and LM alone. When soil is well aerated, the oxidation, i.e. nitrification, of available N dominates and NO is the most common gas emitted from soil instead of $N_2O$ (Davidson et al., 2000). As the rice-growing season progresses, N is taken up by the plant or otherwise leached (Cassman et al., 1998), resulting in low $N_2O$ emissions (Xing et al., 2002; Pandey et al., 2014). When taking into account both $CH_4$ and $N_2O$ fluxes, the LED treatments had 7.5 and 4.5 times lower total global warming impacts than CF and MD respectively. In light of the likely increase in straw additions and the practical challenges of drainage, the LED strategy will facilitate the mineralising of C that can have long-term effects on emission and is likely to be a better option than the CF and MD strategy in terms of the total GHG budget.

4.4. GWP, yield-scaled GWP

The area-based GWP of $CH_4$ and $N_2O$ emissions in the CF treatment were 21,705 and 637 mg per soil column (corresponding to 14,094 and 414 kg $CO_2$ equiv. ha$^{-1}$) respectively. This is in agreement with the meta-analysis by Feng et al. (2013) who report an average GWP of $CH_4$ and $N_2O$ emissions of 14,331 and 699 kg $CO_2$ equiv. ha$^{-1}$ respectively. The long early-season drainage treatment showed very high mitigation potential, reducing the area-based GWP of $CH_4$ and $N_2O$ to 2203 and 755 mg per soil column (corresponding to 1431 kg and 490 kg $CO_2$ equiv. $ha^{-1}$) respectively. A more relevant measure might be yield-scaled emissions, linking GHG emissions directly to agricultural production (Pathak et al., 2010; Van Groenigen et al., 2010; Linquist et al., 2012a, 2012b; Venterea et al., 2011). In the present study, plant growth was clearly affected by ED treatments, most markedly by larger vegetative biomass and plant vigour, which may have been due to better root growth and development (Table 2). Despite the inclusion of conditions that could trigger high $CH_4$ emissions, the LED treatment LE + LM showed potential to reduce yield-scaled GWP by 87% with only 0.43 kg $CO_2$ equiv. $kg^{-1}$ rice grain (430 kg $CO_2$ equiv. $Mg^{-1}$). This yield-scaled GWP from the LE + LM treatment was 62% lower than the yield-scaled GWP (1146.3 kg $CO_2$ equiv. $Mg^{-1}$) reported by Pathak et al. (2010) and 34% less than the yield-scaled GWP from the global rice production (657 kg $CO_2$ equiv. $Mg^{-1}$) reported by Linquist et al. (2012a, 2012b).

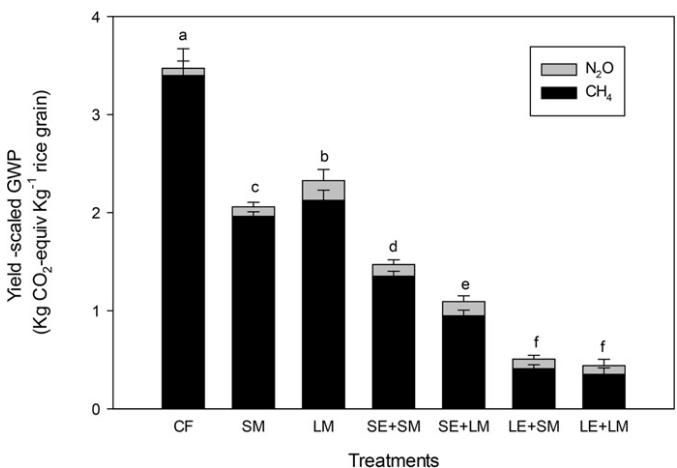

Fig. 6. Yield-scaled GWPs as affected by different alternate water management regimes. CF = Continuous flooding, SM = Short Mid-season drainage, LM = Long Mid-season drainage, SE = Short early-season drainage, LE = Long early-season drainage. Error bars indicate 1 S.E.M. (n = 3). Different letters indicate significance (p < 0.05) of treatments of rice production (small letters).

However, the present study was limited by controlled laboratory conditions and relatively small soil columns and therefore yield and yield-scaled GHG data are not directly comparable to field conditions. Given the very high importance of yield and yield-scaled data to farmers if they are to adopt new mitigation options, yield and yield-scaled emission are represented here. Despite being a highly controlled growth chamber study, these results are considered to be highly relevant for future field studies in different geographical regions. First, the measured emissions (area as well as yield-scaled) were found to be within the levels found in field studies, as reported in recent meta-analyses (Yan et al., 2009; Feng et al., 2013). Second, conditions were chosen that closely mimic those in paddy fields. Reports from rice production systems indicate that differences in $CH_4$ emissions are the main contributors to significant differences in yield-scaled GWP. Several factors can significantly affect $CH_4$ emissions from paddy fields, mainly soil temperature, crop residue management and water management. This study carefully mimics the tropical rice field conditions in climate chambers where soil temperature closely resembles topical soil temperatures along with a large input of rice straw (10 t $ha^{-1}$). Similarly, the N fertilisation dose was carefully selected following the guidelines from the meta-analysis by Feng et al. (2013), which showed that the largest reduction in yield-scaled GWP occurred at the rate of 150–200 kg N $ha^{-1}$. Some previous studies (Towprayoon et al., 2005; Li et al., 2011; Xu et al., 2015) report that MD and intermittent irrigation results in a reduction in rice yield, indicating the importance of considering drainage regimes both for the impacts on rice yield and GHG emissions, which guided the design of the current study. In future field studies, it might also be of interest to extend the calculation of net GWP to include the soil carbon balance (Mosier et al., 2006).

### 4.5. Farmers' perspectives, limitations and the adoption of applicable alternative water management regimes

Rice production uses approximately 40% of the world's irrigation water. Close to one third of these areas experience water shortages, necessitating water-saving strategies without compromising yield. Mid-season drainage alone can reduce emissions by up to one third compared to CF, but practices such as alternate wetting and drying (AWD) have shown greater water-saving and emission reduction potential (Sander et al., 2015). However, to practise AWD, farmers must first be able to allow their fields to dry, and then must have a reliable source of water to rewet their fields as soon as it is needed and repeat this in a more or less continuous cycle. Therefore in order to implement these practices, farmers need reliable control over irrigation water and usually also require small, well-levelled fields to avoid pockets that dry excessively in the distant part of the field that would impact rice yields with repeated drying cycles. In many developing Asian countries, full-scale AWD is therefore often not feasible because farmers have limited technical ability to sufficiently drain and re-flood their fields during the rainy season. In the dry season, farmers who rely on surface irrigation systems have a tendency to be reluctant to interrupt irrigation when water is available because of uncertainties around water availability when it is needed. In some of these locations, early-season drainage plus mid-season drainage could be an effective means of reducing methane emissions, since a delay to the start of flooding at the start and a single long mid-season drawdown may still be feasible. However, emission reductions alone do not motivate the adoption of these water management techniques since they do not directly benefit farmers. In areas where farmers receive water through gravity-driven irrigation, these farmers rarely benefit financially from reducing their water use because they do not pay for the quantity of water they use. In contrast, many farmers who rely on pump-driven irrigation do directly benefit from saving water, providing a potential incentive to reduce the duration of flooding. To date the low adoption of mitigation strategies, e.g. AWD, indicates the importance of incentives to increase their adoption. Incentives such as carbon payments as a form of compensation for reducing emissions compared to traditional practice and subsidised agricultural inputs, e.g. good quality seeds, fertilisers etc. would encourage farmers to test new mitigation techniques. No significant reduction in yield was found between the conventional CF and LED drainage regimes, meaning that even without incentives the LED regime appears to be economically competitive with lower input costs, e.g. water savings and equivalent yield. However, it is recognised that the high potential of ED treatments may not be fully reached under field conditions as our plant growth chamber experiment was conducted under near-optimal conditions. In particular, it is difficult to maintain homogenous water level conditions and homogeneous levels of applied residue in the field. Therefore our results should be validated in field experiments in different geographical regions with differing plant growth conditions.

### 5. Conclusions

Our results suggest a strong potential for early-season drainage to reduce the total GHG budget from rice paddy systems. Although the results should be validated in the field under realistic conditions, the LED treatments appeared to effectively mitigate seasonal $CH_4$ emissions relative to conventionally managed CF and MD water regimes while still maintaining grain yield. They also significantly reduced net yield-scaled GHG emissions. No significant difference was found between the two LED treatments (LE + SM and LE + LM), confirming the fact that the duration of mid-season drainage is not important when the duration of early-season drainage is sufficiently long. These results also strongly suggest that LED and SED can facilitate a decrease in net GWP from rice paddy fields. Therefore, short and long early-season drainage treatments are proposed in addition to one mid-season drainage episode as an effective mitigation measure. In regions where farmers have limited technical ability to drain their fields on a regular basis and face uncertainty in water availability, too much wet season water or uneven fields, the full-scale practice of AWD is not feasible. Simple alternate drainage regimes such as ED + MD might be an effective low-tech option for those small-scale farmers to reduce GHG emissions and save water while maintaining yield.

### Acknowledgements

This work has been conducted as part of a Ph.D. fellowship project supported by the Agricultural Transformation by Innovation (AGTRAIN), Erasmus Mundus Joint Doctorate Programme, funded by the EACEA (Education, Audiovisual and Culture Executive Agency) of the European Commission (agreement no. 2013-008). This work was further supported by the Climate and Clean Air Coalition (CCAC) and the CGIAR Research Program on Climate Change, Agriculture and Food Security (CCAFS), which is carried out with support from CGIAR Fund Donors and through bilateral funding agreements. For details please visit https://